\documentclass[twocolumn,showpacs,preprintnumbers,amsmath,amssymb]{revtex4}

\usepackage{graphicx}
\usepackage{dcolumn}
\usepackage{bm}

\begin{document}


\title{Dephasing in (Ga,Mn)As nanowires and rings}

\author{K. Wagner}
\email{konrad.wagner@physik.uni-regensburg.de}
\author{D. Neumaier}
\author{M. Reinwald}
\author{W. Wegscheider}
\author{D. Weiss}
\affiliation{Institut f\"ur Experimentelle und Angewandte Physik,
Universit\"at Regensburg, Regensburg, Germany}

\date{\today}

\begin{abstract}
To understand quantum mechanical transport in ferromagnetic
semiconductor the knowledge of basic material properties like phase
coherence length and corresponding dephasing mechanism are
indispensable ingredients. The lack of observable quantum phenomena
prevented experimental access to these quantities so far. Here we
report about the observations of universal conductance fluctuations
in ferromagnetic (Ga,Mn)As. The analysis of the length and
temperature dependence of the fluctuations reveals a $T^{-1}$
dependence of the dephasing time.
\end{abstract}


\maketitle

The discovery of the ferromagnetic III-V semiconductor materials
(In,Mn)As \cite{Ohno} and (Ga,Mn)As \cite{Ohno2} has generated a lot
of interest as these materials combine ferromagnetic properties,
typical for metals, with the versatility of semiconductors (for a
review see e.g. \cite{Ohno3,Ohno4,Ohno5,Dietl,MacDonald}). This
allows, e.g., to control ferromagnetism by electric fields thus
opening new prospects for application and fundamental research
\cite{Ohno6}. The Mn atoms in the III-V host are not only
responsible for the ferromagnetism but also act as acceptors such
that, at sufficiently high Mn-concentration (Ga,Mn)As is a
degenerate p-type semiconductor \cite{Oiwa}. The
ferromagnetic order of the Mn magnetic moments is
mediated by holes via the Ruderman-Kittel-Kasuya-Yosida (RKKY)
interaction \cite{Dietl2}. The quest to increase the Curie
temperature $T_\mathrm{C}$ in (Ga,Mn)As towards room temperature has
led to a thorough investigation of the material properties (see, e.g.
\cite{Jungwirth} and references therein). By annealing (Ga,Mn)As sheets or by incorporating
them into sophisticated layered arrangements the Curie temperature
was increased up to 173~K \cite{Edmonds,Wang} and 250~K with Mn
$\delta$ doping \cite{Nazmul}, respectively. Despite the high
crystalline quality of the material (Ga,Mn)As is a quite disordered
conductor on the verge of the metal-insulator transition (MIT). For
Mn concentrations on the metallic side of the MIT the typical mean
free path of the holes are a few lattice constants. Hence quantum
effects like Shubnikov-de Haas or Aharonov-Bohm (AB) oscillations or
universal conductance fluctuations (UCF) were not yet reported.
Hence the phase coherence length $L_{\phi}$  and the corresponding
dephasing mechanisms which govern quantum mechanical interference
phenomena in ferromagnetic semiconductors are not known yet. Below
we report the observation of UCF in
nanoscale (Ga,Mn)As wires at very low temperatures and probe
currents. From an analysis of the temperature and length dependence
of the UCF's information about phase coherence and dephasing in
(Ga,Mn)As is obtained.

\begin{figure}
\includegraphics[width=0.9\columnwidth]{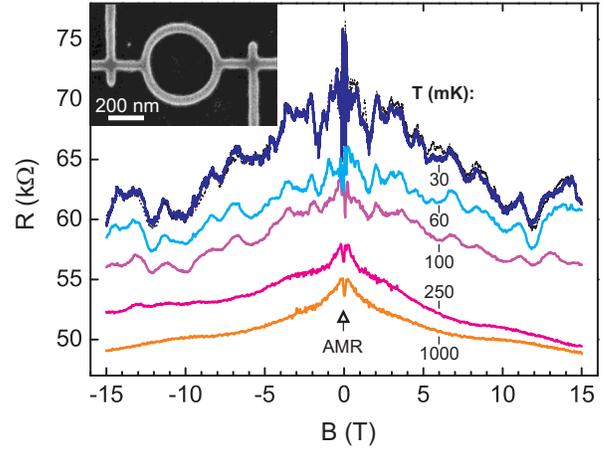}
\caption{\label{Fig1}(color online). Magnetoresistance of a
(Ga,Mn)As-ring with a diameter of 400~nm and a ring width of 40~nm.
The inset displays a top view of the ring. The temperature was
varied from 1~K (bottom trace) to 30~mK (top trace) and the current
through the device was set to 100~pA. To demonstrate the
reproducibility of the resistance oscillations observed below
$\sim200$~mK the 30~mK trace is shown for an up-(blue line) and
down-sweep (dashed black line) of the $B$-field.}
\end{figure}

For our experiments, nanoscale wire and ring samples were fabricated from two different
wafers grown by low temperature molecular beam epitaxy on semi
insulating GaAs (100) substrates \cite{Reinwald}. The (Ga,Mn)As
layers were 50~nm thick with a Mn concentration of $\sim2$\%. From
transport experiments $T_\mathrm{C}$ was
evaluated to be $\sim60$~K. The sheet resistance $R_{\square}$ is
1.8~k$\Omega$ at 1~K corresponding to a dimensionless
conductance
$g=\frac{1}{R_{\square}}\left(\frac{e^{2}}{h}\right)^{-1}\approx14\gg1$. The
hole carrier concentration $p$, extracted from the high-field Hall resistance $R_{xy}$ at 90~mK is
$\sim1.8\times10^{20}~\mathrm{cm}^{-3}$ \footnote{We obtained $p$ by
probing the best fit to $R_{xy} = R_{0}B/t + R_{s}M/t$
\cite{Omiya} and extracting $p$ from the ordinary Hall constant
$R_{0}=1/pe$ above $B>0.2$~T where the magnetization $M$ is
saturated. Here the magnetic field $B$ is perpendicular to the
sample's plane, $t$ is the layer thickness, and $R_{s}$ is the
anomalous Hall term. $R_{s}$ depends on $R_{sheet}$ as $R_{s}\propto
R_{sheet}$ or $R_{s}\propto R_{sheet}^2$. Our fits showed clearly a
$R_{s}\propto R_{sheet}$ dependence for the material investigated.}.
Within a 6 band $k\cdot p$ model this corresponds to a Fermi energy
of 130~meV \cite{Dietl3}. Using the joint effective mass of
0.61$m_{0}$ for heavy and light hole bands in GaAs we obtain a
momentum relaxation time $\tau_{p}=1.3$~fs. Here, $m_{0}$ is the
free electron mass. Thus the low temperature diffusion constant
$D=\frac{1}{3}v_{F}^{2}\tau_{p}=4.8\times10^{-5}~\mathrm{m^{2}/s}$
where $v_{F}$ is the average Fermi velocity of heavy and light
holes.

\begin{figure}
\includegraphics[width=0.9\columnwidth]{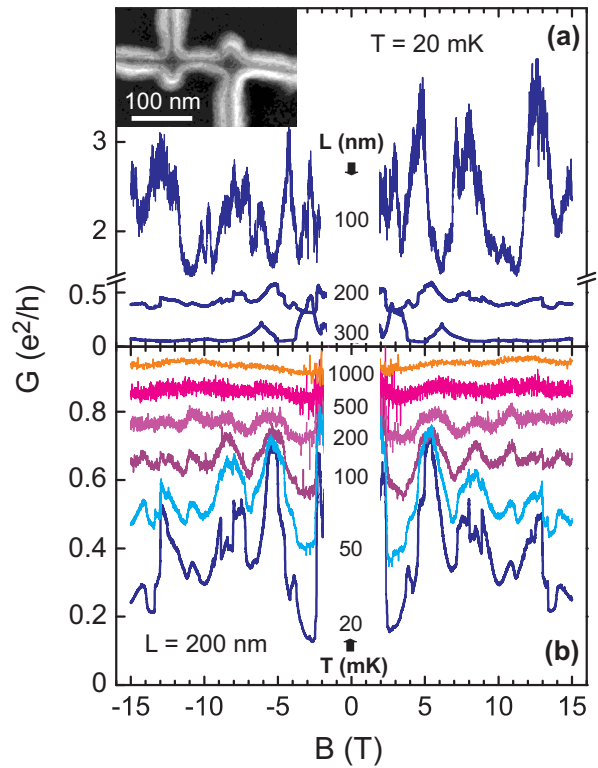}
\caption{\label{Fig2}(color online). (a) Conductance fluctuations
for three wires of different length $L$. For the shortest wire the
amplitude of the conductance fluctuations is about $e^{2}/h$,
expected for conductors with all spatial dimensions smaller or
comparable than $L_{\phi}$. The inset shows an electron micrograph
of a 20 nm wide wire with a potential probe separation of
$\sim100$~nm (the 100~nm wire). (b) $G$ vs. $B$ of the 200~nm wire
for different temperatures between 20~mK and 1~K. In both
experiments the current through the wire was 10~pA (exception: 30~pA
at 1~K). The conductance traces in the low $B$-regime have been
omitted in both graphs.}
\end{figure}

For transport experiments (Ga,Mn)As layers were first pre-patterned
with a Hall-bar structure using  standard optical lithography and
chemical dry etching. Au contacts were made by standard lift-off
method after brief in-situ ion beam etching of the surface to remove
the native oxide layer. To define wires and rings we used electron
beam lithography (a LEO Supra 35 SEM and nanonic pattern generator)
and reactive ion etching. Wires with lengths of 100~nm to 15~$\mu$m
and a nominal width as small as 20~nm (see inset of Fig.~\ref{Fig2})
were fabricated. To probe the AB effect we also fabricated rings
with diameters of 100~nm to 800~nm and widths of 20~nm to 40~nm. An example of a
ring with 400~nm diameter is shown as inset in Fig.~\ref{Fig1}. The
magnetotransport properties were measured in a top-loading dilution
refrigerator with a base temperature of 15~mK and magnetic
fields up to 19~T (Oxford Kelvinox TLM Dilution Refrigerator with
Femtopower Thermometry using $\mathrm{RuO_{2}}$ sensors calibrated
down to 20~mK). To avoid heating we used a low frequency (17~Hz) and
low current (10~pA to 100~pA) four probe lock-in technique. Low
noise preamplifiers and careful wiring and filtering were crucial to
observe quantum interference effects in this material.

In an attempt to observe AB oscillation we fabricated (Ga,Mn)As
rings and measured their resistance as a function of a perpendicular
magnetic field $B$ and temperature $T$. Corresponding data are shown in
Fig.~\ref{Fig1}. At high temperatures of 1~K and 250~mK the
magnetoresistance exhibits the behavior characteristic of (Ga,Mn)As
layers: The resistance first increases up to a field of $\sim120$~mT and
then decreases over the entire investigated field range. The
low-field increase is ascribed to the anisotropic magnetoresistance
effect (AMR) \cite{Baxter} which flattens once the magnetization
is saturated. The
negative magnetoresistance is commonly assigned to decreasing
magnetic disorder \cite{Nagaev} but by others also seen as a
signature of weak localization \cite{Matsukura}. At temperatures
below $\sim200$~mK reproducible resistance fluctuations emerge but
no AB oscillations are seen. The latter is expected if the ring
circumference exceeds the phase coherence length $L_{\phi}$
significantly. The amplitude of the resistance fluctuations
depends strongly on temperature and current. Hence the
current through the ring was limited to 100~pA here. The resistance
fluctuations are especially pronounced in the low field interval
dominated by the AMR. These fluctuations are out of the scope of the
discussion below. As the magnetization in this low field regime is
expected to be inhomogeneous an analysis of these fluctuations is
complicated and might require a detailed knowledge of the local
magnetization. Hence we focus on the high field fluctuations which
occur in a $B$-regime of homogeneous magnetization below. Such a
reproducible (within one cooling cycle) stochastic
magnetoresistance pattern for a specific sample, also denoted as
'magnetofingerprints', is known as UCF \cite{Lee}.

To investigate the fluctuations we measured the conductance
of (Ga,Mn)As wires with lengths between 100~nm and 15~$\mu$m and a width of 20~nm.
Since the Hall resistance contribution
$R_{xy}$ stemming from both anomalous and regular Hall effect is by
a factor of $\sim20$ smaller than the longitudinal resistance
$R_{xx}$, the conductance $G$ is simply obtained by inversion, $G=1/R_{xx}$.
The magnetic field dependence of
$G$ of wires with 100~nm, 200~nm and 300~nm lengths,
measured at 20~mK, is shown in Fig.~\ref{Fig2}a. With decreasing
lengths of the wires not only the conductance \footnote{We have
checked that we are still in a regime where the conductance of the
wires scales as $1/L$ whith $L$ the length of a wire. Sample to
sample fluctuations of the actual width in our extra narrow wires
cause deviations from this scaling if, as in Fig.~\ref{Fig2}a,
different samples with the same nominal width are compared.} but
also the amplitude of the fluctuations increases. At
20~mK and for the shortest wire of 100~nm length the amplitude of
the fluctuations is of order $e^{2}/h$. The disappearance of the
fluctuations for temperatures above 200~mK is
demonstrated for the 200~nm wire in Fig.~\ref{Fig2}b.

\begin{figure}
\includegraphics[width=0.9\columnwidth]{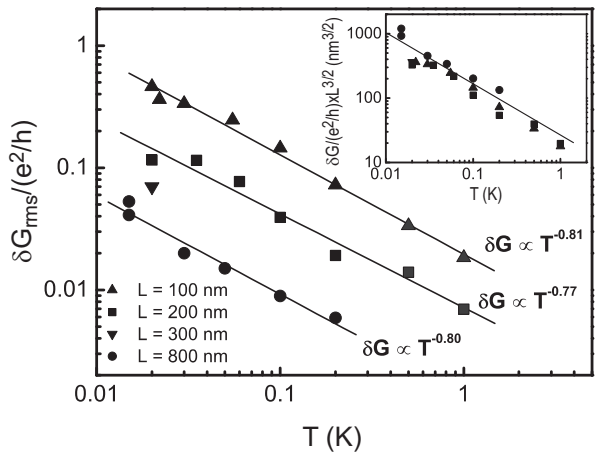}
\caption{\label{Fig3} $\delta G$ in units of $e^{2}/h$ for wires of
length 100~nm, 200~nm and 800~nm. The measurements were carried out
using currents of 10-500~pA. The maximum current at a given
temperature was adjusted such that the fluctuation amplitude $\delta
G$ was not affected. Inset: $\delta
GL^{3/2}/(e^{2}/h)=L_{\phi}^{3/2}$ vs $T$.}
\end{figure}

UCF result from correlations between
different transmission paths through a disordered sample
(\cite{Washburn} and references therein). If the wire is smaller than $L_{\phi}$ in all three dimensions the
fluctuation amplitude $\delta G=\sqrt{\langle (G-\langle
G\rangle)^{2}\rangle}\approx e^{2}/h$, where the bracket
$\langle...\rangle$ denotes averaging over $B$. Hence the data
in Fig.~\ref{Fig2}a suggest that the phase coherence
length is of order 100~nm for the lowest temperatures since $\delta
G$ is of order $e^{2}/h$ for the 100~nm wire. Once the dimensions of
the conductor exceed the phase coherence length
$L_{\phi}=\sqrt{D\tau_{\phi}}$, with the coherence time
$\tau_{\phi}$, $\delta G$ gets suppressed \cite{Lee,Washburn}. To
elucidate the conductance fluctuations further we study the
fluctuation amplitude as a function of wire length and temperature.
Corresponding data for wires with $L=100$~nm, 200~nm and
800~nm are displayed in Fig.~\ref{Fig3}. The conductance amplitude $\delta G$ of all wires shows a
power law dependence with an exponent close to $-3/4$. Carrying out
the experiments at finite $T$ another length scale, the
thermal length $L_{T}$ enters the stage. This length sets the scale over
which thermal smearing causes dephasing. If the width $w$ and the
thickness $t$ of the wires $t,w<L_{\phi}<L_{T}$ the dependence of
$\delta G$ on $L_{\phi}$ is given by \cite{Lee}
\begin{equation}\label{dg}
    \delta
    G=\frac{e^{2}}{h}\left[\frac{L_{\phi}}{L}\right]^{3/2}.
\end{equation}
This formula is valid for one-dimensional (1D) conductors.
Since we expect $L_{\phi}$ on the order of 100~nm, width and
thickness of our wires ($w\approx20$~nm, $t\approx50$~nm) are smaller than
$L_{\phi}$ and the wires can be considered as 1D-conductors. Using
the diffusion constant $D$ estimated above we obtain for
$L_{T}\approx150$~nm at 15~mK. This means that the condition
$t,w<L_{\phi}<L_{T}$ is realized in the (Ga,Mn)As wires, at least
for low $T$. If the data in Fig.~\ref{Fig3} are
multiplied by $L^{3/2}$, all data points collapse on one line,
shown in the inset of Fig.~\ref{Fig3}. Hence the scaling of $\delta
G$ with $L$ is correctly described by Eq.~(\ref{dg}).
The temperature dependence of $L_{\phi}$ can then be
directly read off the inset in Fig.~\ref{Fig3} since, according to
Eq.~(\ref{dg}), the ordinate is given by $L_{\phi}^{3/2}$.
Extrapolating to 10~mK a phase coherence length of $\sim100$~nm is
obtained.

The data in Fig.~\ref{Fig3} imply that the temperature dependence of
the phase coherence length is $\propto T^{-1/2}$ or, put in terms of
the coherence time, $\tau_{\phi}\propto T^{-1}$. Here we have
assumed that the $T$-dependence of $\delta G$ is, accordant
with the data of Fig.~\ref{Fig3}, given by $T^{-3/4}$. We note that
the $T$-dependence of $L_{\phi}$ remains unchanged even in the
regime where $L_{T}<L_{\phi}<L$. There, instead of Eq.~(\ref{dg})
$\delta G=(L_{T}/L)(L_{\phi}/L)^{1/2}e^{2}/h$ holds \cite{Lee}.
Since $L_{T}\propto T^{-1/2}$ and $\delta G\propto T^{-3/4}$, again
$L_{\phi}\propto T^{-1/2}$ results.

The $T$-dependence of $L_{\phi}$ contains information about
the relevant dephasing mechanism and usually obeys a power law of
the form $L_{\phi}\propto T^{-\alpha}$ where $\alpha$ depends on the
dominating phase breaking mechanism and the dimensionality $d$
($d=1$ in our case). For $\alpha=-1/2$ no dephasing mechanism is readily
available. Dephasing by electron-phonon scattering in 1D is usually described by $L_{\phi}\propto T^{-1}$ to
$T^{-2}$ \cite{Bird}. At low $T$ and reduced dimensions the
small energy transfer Nyquist electron-electron interaction is a
common source of dephasing \cite{Altshuler}. The corresponding
$L_{\phi}\propto T^{-1/3}$ dependence does, however, not describe
our experimental result. A possible candidate for dephasing is
critical electron-electron scattering described
for strongly disordered metals in the vicinity of the MIT
\cite{Dai}. The corresponding inelastic scattering time was
calculated to be $\propto T^{-1}$ but the calculations were done for
bulk systems \cite{Belitz}. One should keep in mind, though, that in
our system the charge is carried by holes. In addition the material is ferromagnetic. It is
interesting to note here that - unlike for mesoscopic conductors
containing magnetic impurities \cite{Pierre} - we do not observe
saturation of coherence time at low temperatures.

\begin{figure}
\includegraphics[width=0.9\columnwidth]{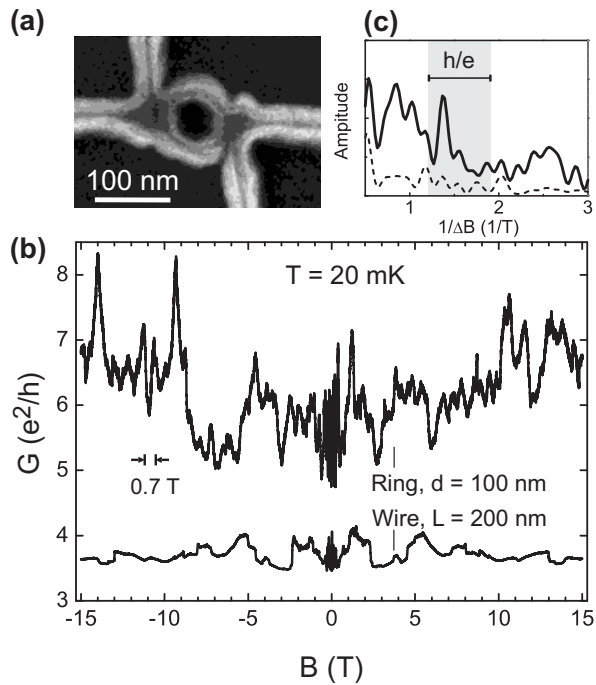}
\caption{\label{Fig4}(a) Electron micrograph of a (Ga,Mn)As ring
sample with a diameter of $\sim100$~nm. (b) Comparison of the
magnetoconductance trace of the ring sample with the conductance of
a wire of comparable length and 20~nm width. (c) Corresponding FFT
taken from the conductance of ring and wire. The region where AB
oscillations are expected is high-lighted.}
\end{figure}

As our analysis of UCFs suggests a phase coherence length of order 100~nm at mK-temperatures, the AB effect should be observable in sufficiently small (Ga,Mn)As rings. A corresponding experimental result is displayed in Fig.~\ref{Fig4}, an electron micrograph of the investigated ring is shown in Fig.~\ref{Fig4}a. From the micrograph we estimate an electrically active inner diameter $2R_{i}$ of only 80~nm, an outer diameter $2R_{o}$ of 100~nm. One half of the average ring perimeter is hence only slightly longer than $L_{\phi}$. The ring's conductance is compared to the conductance of a wire of 20~nm width and 200~nm length. From the ring's dimension we expect AB oscillations with a period $\Delta B$ between $h/e\pi R^{2}_{o}=0.53$~T and $h/e\pi R^{2}_{i}=0.82$~T. Clear conductance oscillations in this range, absent for the wire geometry, are superimposed on the ring's conductance in Fig.~\ref{Fig4}b. A bar of $\Delta B=0.7$~T length is displayed as a guide for the eye there. A Fourier transform of both, ring and wire conductance is compared in Fig.~\ref{Fig4}c. A peak at $1/\Delta B\sim1.4~\mathrm{T}^{-1}$ appears in the Fourier transform lying within the window expected for AB oscillations. While clear AB oscillation are observable in our smallest ring, the analysis is complicated by two factors: (i) superimposed upon the oscillations are conductance fluctuations with a similar magnetic field scale as the corresponding areas enclosed by trajectories are similar \cite{Washburn2}. Hence fluctuations and oscillations are hard to separate. (ii) The finite width of the ring causes an uncertainty in the expected $\Delta B$ of the oscillations. Trajectories traveling along the outer side of the ring might - in contrast to paths along the inner perimeter - contribute less to the AB effect.

By resolving both universal conductance fluctuations and Aharonov-Bohm effect we have shown clear phase coherent effects in ferromagnetic (Ga,Mn)As nanostructures. The phase coherence length in the material investigated was $\sim100$~nm at 10~mK and showed a $T^{-1/2}$ temperature dependence. This means that the relevant coherence time follows a $T^{-1}$ law.

Acknowledgements

We thank T. Dietl, J. Fabian, and C. Strunk for valuable discussions
and the German Science Foundation (DFG) for their support via the
Research Unit FOR370 and the Collaborative Research Grant SFB689.

\bibliography{paper}

\end{document}